\documentclass{amsart}
\usepackage{a4wide}
\usepackage{amsfonts}
\usepackage{psfrag}
\usepackage{color}
\usepackage[all]{xypic}
\usepackage{hyperref}
\usepackage{amsmath}
\usepackage{amssymb}
\usepackage{amsthm}
\usepackage{graphicx}
\usepackage{mathtools}

\newtheorem{theorem}{Theorem}[section]

\theoremstyle{definition}
\newtheorem{definition}[theorem]{Definition}
\newtheorem{remark}[theorem]{Remark}

\numberwithin{equation}{section}
\numberwithin{equation}{section}
\numberwithin{equation}{section}

\begin{document}
\title{A Data-driven Market Simulator 
\\for Small Data Environments}

\author{Hans B\"{u}hler}
\address{J.P. Morgan, London}
\email{hans.buehler@jpmorgan.com}

\author{Blanka Horvath}
\address{
Department of Mathematics, King's College London and The Alan Turing Institute}
\email{blanka.horvath@kcl.ac.uk and b.horvath@imperial.ac.uk}

\author{Terry Lyons}
\address{Mathematical Institute, University of Oxford and The Alan Turing Institute}
\email{terry.lyons@maths.ox.ac.uk}

\author{Imanol Perez Arribas}
\address{Mathematical Institute, University of Oxford and The Alan Turing Institute}
\email{imanol.perez@maths.ox.ac.uk}

\author{Ben Wood}
\address{J.P. Morgan, London}
\email{ben.wood@jpmorgan.com}

\date{\today}
\thanks{Opinions expressed in this paper are those of the authors, and do not necessarily reflect the view of JP Morgan.
%\\B.H.is grateful for the opportunity of a research oriented quantitative internship at JP Morgan Chase, in course of which this research was carried out.
The algorithm developed in this paper is available in the Github repository provided with this paper \href{https://github.com/imanolperez/market_simulator}{Github:Marketsimulator}
}

\maketitle

\begin{abstract}
Neural network based data-driven market simulation unveils a new and flexible way of modelling financial time series, without imposing assumptions on the underlying stochastic dynamics. Though in this sense \emph{generative market simulation} is model-free, the concrete modelling choices are nevertheless decisive for the features of the simulated paths. 
We give a brief overview of currently used generative modelling approaches and performance evaluation metrics for financial time series, and address some of the challenges to achieve good results in the latter. We also contrast some classical approaches of market simulation with simulation based on generative modelling and highlight some advantages and pitfalls of the new approach.  
While most generative models tend to rely on large amounts of training data, we present here a generative model that works reliably even in environments where the amount of available training data is notoriously small. Furthermore, we show how a \emph{rough paths} perspective combined with a parsimonious Variational Autoencoder framework provides a powerful way for encoding and evaluating financial time series in such environments where available training data is scarce. Finally, we also propose a suitable performance evaluation metric for financial time series and discuss some connections of our Market Generator to deep hedging.
\end{abstract}

\tableofcontents

\section{Introduction}\label{sec:Introduction}
Recent advances of deep learning applied to quantitative finance \cite{DLV219,BGTW18,CKT20,ContHamida,PHL2, DLV19,RufWang19, Zhang} have demonstrated the potential prowess of deep learning algorithms in the context of pricing and hedging derivatives. While the neural network based hedging engine presented in \cite{BGTW18} rendered a convincing hedging performance when the market scenarios provided for training came from classical stochastic models (such as the Black-Scholes and Heston models including various forms of market frictions), the approach is in fact inherently model agnostic. The collection of sample paths provided to the network in the training phase enable the latter to generate optimal response strategies with respect to market distributions resembling the ones presented during the training phase. This and similar model-agnostic neural network based financial applications drove interest towards such market scenario generators that are flexible and highly realistic, ideally even model-free and directly data driven. Indeed, the accurate modelling and efficient numerical simulation of stochastic market movements and financial time series has been a central theme throughout the past decades of financial modelling. In practice however, even an overly complex model family does not necessarily include the target function or the true data-generating process (or a close approximation thereof) when the latter evolves over time. In this work we take a leap beyond classical stochastic models and present a statistically driven market simulator based on generative modelling:
Lately, tremendous progress has been made in training neural networks as powerful function approximators through backpropagation. This progress has brought about frameworks which can use backpropagation-based function approximators to build generative models. Such models are based on the idea of transforming samples of latent variables to data samples via differentiable functions, which are approximated by a neural network.
Indeed, the emergence of generative modelling techniques in various machine learning applications opened up new horizons for even more flexible and directly data-driven simulations of market paths, however these possibilities come along with a new set of challenges which brings new unexplored questions to light that we endeavour to identify and address in this paper. The contributions presented in this work are the following: We develop a powerful, flexible, non-parametric generative model for financial time series based on signatures of paths, which is capable of efficiently operating in an environment where a very low amount of training data is available. We demonstrate the prowess of this algorithm and its ability to produce realistic synthetic data to a high precision---that can be conditioned on various market indicators---in various numerical examples on historical-, as well as on simulated data.\\  
The paper is organised as follows:  We motivate our work by outlining in Section \ref{sec:PracticalApp} how
potential applications of generative modelling based market simulation (market generators) can go beyond the currently widespread standard applications of numerical simulations of market paths. These applications incldude data anonymisation techniques, or the identification of relevant properties (anomalies and outliers) of available data, as well as sophisticated backtesting and risk management strategies. Section \ref{sec:PracticalApp} contains a brief summary of such applications, for a more detailled study see \cite{KonSchHo2020}.  In Section \ref{sec:challenges} we contrast the main differences of traditional path generation of stochastic processes with the generative approach and discuss new challenges that arise if numerical time series are simulated by generative models rather than traditional means. We also point out how some of the traditionally common indicators describing if a financial time series is a realistic reflection of market paths may--- for generative models---not be fully sufficient as performance evaluation metrics. 
To address the aforementioned new set of challenges we develop an approach that uses the \emph{signatures} of historical path segments as input data. The concept of signatures was coined in the context of \emph{Rough Paths} (see \cite{FrizHairer,L14,LTL07}) and provides an efficient and parsimonious tool kit to encode the most essential information contained in (continuously observed or discretely sampled) market paths.  
 \\
In Section \ref{sec:MainResults} we present our main results and modelling setup. Section \ref{sec:Methodology}  gives a brief overview of our general methodology summarized in five main steps and Section \ref{sec:BackgMotMethodology} provides more details on the modelling choices in each of the steps.
To demonstrate the efficiency of our proposed approach numerically in Section \ref{sec:NumericalExperiments}, we generate synthetic market scenarios using variational autoencoders both in the signature-based setting and in the standard returns-based setting. The superiority of the signature based framework is displayed both from a computational and from a theoretical point of view. In order to address the non-stationary nature of financial markets, in our framework the generated new market scenarios can be conditioned on various market indicators. The latter conditioning can then also be used to build sample paths of arbitrary length (see Section \ref{sec:Methodology} (\emph{Step 3}) and Section \ref{sec:NumericalExperiments}). It is apparent in the majority of (deep) neural network learning based pricing, hedging, forecasting or even generative algorithms that these applications often heavily rely on the availability of large training datasets, which are not always readily available. The particular achievement presented in this work is to remove this limitation: We demonstrate the ability of our approach to generate new samples from a particularly small observable dataset to a high precision, which can in turn be conditioned on specific market indicators and conditions and used to feed other applications with the necessary amount of training samples. The algorithm developed in this paper is available in the Github repository provided with this paper \href{https://github.com/imanolperez/market_simulator}{Github:Marketsimulator}. 

%%%%%%%%%%%%%%%%%%%%%%%%%%%%%%%%%%%%%%%%%%%%%%%%%%

\subsection*{Generative Modelling and Market Generators} The emergence of DNN-based financial applications is one of the driving factors that directed the interest towards highly realistic market simulators:
A key factor for training these deep networks to a sufficient accuracy is the the availability of sufficiently large, representative training datasets.
In the example of deep hedging it is easy to see that the quality and quantitiy of available training data impacts the hedging performance, since 
unrealistic training data can lead to large losses (resulting from wrongly hedged positions) when the algorithms are subsequently applied to real life data. Though the surge of machine learning came hand in hand with the explosion of available data, in more situations than not, the amount of available training data is insufficient rather than large or the data available for training is not representative of the market, and numerical generation of additional data is necessary. Similarly, failing to train on unseen rare events may leave the application without an adequate response strategy in case the market moves to unexplored territories.
This gave rise to an increase in interest for realistic numerical simulation of financial markets.\\
Deep neural networks provide a powerful tool to approximate complex distributions and this capacity, together with the increases in available computational power and speed has opened new horizons in all areas of modelling, including market simulation. The fundament of the prowess is laid by the universal approximation properties of neural networks \cite{Hornik89, Hornik91}, which establish that any function or distribution with sufficient regularity can be approximated by a sufficiently large neural network.  \emph{Generative models} capture probability distributions by approximating these via neural networks from learned samples, from which new synthetic data samples can be drawn. The generative model is trained such that it is representative the underlying distribution of the given dataset with respect to some loss function referred to as \emph{performance evaluation metric}.  Generative modelling originates from more traditional applications of machine learning and the adaptation of these techniques to financial setting has its bespoke challenges. Identifying and addressing a these challenges is one of the contributions of this work.  %In this work we present a market generator capable not only of closely mirroring historical price movements but also of capturing the relevant market information, and generating alternative market scenarios conditional on observed market events. \\

\begin{definition}[Market Generator]
We refer to generative models in a financial time series context as Market Generators. That is, to neural networks that are designed to approximate the underlying distribution of an underlying market from a data sample given in form of a time series, so as to generate new data variations of the learned distribution.  
\end{definition}

\noindent Possible applications that call for generative simulation of financial markets include: 

\subsection*{Practical use and applications of Market Generators}\label{sec:PracticalApp}
There are several situations in which it is beneficial to rely on simulated data samples that are statistically indistinguishable from a given original dataset. This section outlines a very brief summary of some 
potential applications of generative modelling based market simulation (market generators) that can go beyond the currently widespread standard applications of numerical simulations of market paths. For a more detailed study see \cite{KonSchHo2020} and \cite{PHL2} for an application to anomaly detection.
\begin{enumerate}
\item[(i)] For data anonymisation: When the available data is confidential, it is desirable to generate anonymised datasets that are representative of the true underlying distribution of the data but cannot be traced back to their origin. Financial data and medical data are often proprietary, or confidential. When testing investment strategies or the effectiveness of a treatment it is imperative not to be able to trace back the datasets to the individual client or patient.
\end{enumerate}
Scenario (i) already showcases some of the essential challenges in this context: Evaluating whether the produced data is representative of the distribution that the observed data stems from, depends on the distributional properties (evaluation metrics) that we control for. Also the level of anonymity achieved by this procedure is a highly interesting question on its own. The study presented in \cite{KonSchHo2020} is devoted to understanding these questions in more detail.An even more challenging situation arises if the size available data sample to train the generative model is very small to begin with:
\begin{enumerate}
\item[(ii)]
Small original training datasets: When there are natural restrictions on the number of available original samples (constraints on number of experiments, legal restrictions on the access to data), the available training data may not be sufficient to train the neural network application at hand, e.g. the hedging engine. Clearly, the more complex the application, the more data samples are needed to train it.  
In such cases also the training of generative models is challenging, due to the low number of available original samples. In this case, the generative neural network faces the same challenges, it has to be trainable on a very low number of data samples. Generative models for sparse data environments therefore need to be as parsimonious and easily trainable as possible. 
\end{enumerate}
Once such a generative network is available, the more complex neural network applications can also be trained, using the intermediate step of the market generator that produces the necessary amount of training samples for the latter, which are statistically indistinguishable from the original data. Further practical applications of market generators include (but are not limited to) the following use cases:
\begin{enumerate}
\item[(iii)] Backtesting: When developing a trading strategy, carrying out a \textit{backtest} to measure how the strategy would perform in a realistic environment is of crucial importance. However, using historical data may result in overfitting of the trading strategy. Having a market simulator capable of generating realistic, independent samples of market paths would allow a more robust backtest less prone to overfitting.
\item[(iv)] Risk management of portfolios -- be it of financial derivatives or trading strategies -- is of utmost importance. A realistic market simulator can be used to generate synthetic paths to estimate various risk metrics, such as Value at Risk (VaR).
\end{enumerate}

\noindent In this paper we present a powerful generative modelling technique for time series generation in the situation that is specifically designed for the type of sparse data environments (ii) described above: Financial time series generation, when the number of available original samples is notoriously small.

\subsection*{Some approaches to numerical data simulation in finance: Classical and new}\label{sec:MarketGenerators}
\begin{itemize}
\item \textbf{Classical modelling:} Numerical simulation of financial time series has a long history in related literature, far preceding the recent surge in financial machine learning research: 
\begin{enumerate}
\item[(i)] Classical approaches include for example classical stochastic market models and autoregressive models and variations of these.
Among their advantages are their tractability and the several decades worth of experience in understanding their mathematical properties. Clearly the advantage thereof is a more straightforward suitability to currently prevalent risk-management frameworks.  Disadvantages may include a relative inflexibility, which can result in modelling inconsistencies.
\item[(ii)] Given the increase in computational powers today, more modelling flexibility can be gained within the realm of ``models with a classical flavour'' by adding complexity and further parameters to the models, taking their weighted averages with weights calibrated to the current market conditions, in the spirit of \cite{DumbergenRogers}. This line of thought opens new avenues that interpolate between model-based and model-free approaches \cite{Obloj}, with their own set of challenges which we will not follow further in this paper. Such an experiment is presented in \cite{DLV19}, similar ideas are further developed in \cite{ViladesSzpruchSiska20}.\\
\end{enumerate}
With several decades worth of understanding the asymptotic and stochastic properties of models and with the steady increase of available computational powers, the  evolution of (classical) models moved toward more and more complex and realistic models. Nevertheless, in recent years we have witnessed several situations where classical models have been challenged by market reality. As one of the consequences, the failure of classical models to fully explain the behaviour of asset prices has led to situations where these models have failed to prescribe the right response strategies. A reality that quantitative investors are painfully aware of. While  classical modelling techniques nevertheless undoubtedly continue to have their merits, ML-based technologies offer a possible alternative to more closely mimic the behaviour of markets in more flexible data-driven ways.\\
\item \textbf{Data-driven modern generative modelling:} 
Approaches to generative modelling are based on the common principle of generating new synthetic data samples whose distribution resembles the distribution of some reference dataset.
One of the most striking differences of modern generative modelling to classical generation of synthetic data is that the explicit knowledge of the underlying data generating distribution is no longer required. Therefore, instead of implementing (an approximaion of) some known distribution or transition density, generative models often approximate the underlying distribution implicitly, by
drawing samples from the latter and comparing their similarity to the original dataset with respect to certain similarity metrics. 
This is in particular true for so called \emph{differential generator networks} where a transformation map is learned through backpropagation from an initial source of randomness to a target distribution.  The two most commonly used generative differential network-based approaches are Variational Autoencoders (VAE) and Generative Adversarial Networks (GAN)  and variations thereof. 
\bigskip\\
A leap away from strictly classical modelling but directly generalising these is the work \cite{NNstochVol17} which in a special case simplifies to the Heston Model and in another case to a GARCH(1,1) model.
Kondratyev and Schwarz propose in \cite{KS19} a restricted Boltzmann Machine (RBM) for time-series generation, controlling for the autocorrelation function and quantiles of the generated time series. Boltzmann Machines are among the first generative models introduced to learn arbitrary distributions over binary vectors. 
Recent contributions to this stream of literature using GANs as generative models include the following: \cite{BKPSL19,CKT20,PHL1,WiBaiWoBu19,WiQuantGan2019,AMWX20}.
To date we are not aware of approaches using VAEs for this purpose.\\
\end{itemize}
%%%%%%%%%%%%%%%%%%%%%%%%%%%%%%%%%%%%%%%%
GANs are unquestionably the most popular differential generator networks, though they are typically data-hungry, and it is often difficult to guarantee their  convergence and stability.  
Variational Autoencoders maximize the likelihood of observing the given (original) data samples under the generated samples \eqref{eq:sequencesgen} and are particularly well-adapted\footnote{\emph{The variational autoencoder approach is elegant, theoretically pleasing and simple to implement} \cite[Chapter 20]{MLBook}.}  to the presented scarce data environment. Recent theoretical connections between autoencoders and latent variable models have indeed brought autoencoders to the forefront of generative modeling (see \cite{MLBook}).  \\
Therefore, in this paper we focus on generative modelling based on Variational Autoencoders (and their conditional refinement) and highlight their prowess in the context of financial time-series generation. More details can be found in Section \ref{sec:BackgMotMethodology} and Appendix \ref{sec:VAE}.\\

\section{Challenges of financial time-series simulation: Classical and new}\label{sec:challenges} 
Currently, many of the available neural network-based generative models originate from static applications (such as image processing) and therefore, several available performance evaluation metrics for generative models have been developed to measure some form of marginal distributions. The incorporation of of a time-series aspect of the data poses additional challenges, one of which is that these static performance evaluation metrics may not always be straightforward to generalise to time series data (see more about this in Secion \ref{sec:CommonEvMetrics}). Simulated financial time series data is commonly addressed by capturing specific universal features of the time series, commonly referred to as \emph{stylized facts}. Below we recall a number of stylised facts that traditional stochastic models are typically aimed to reflect.
Though for classical stochsatic models, these stylized facts are often formulated in terms of the distribution of returns, this returns-based viewpoint (though still interesting) may not be the ideal choice to convey a sufficiently full picture for distributions of synthetic market paths that come from market simulators using generative modelling. Some further issues arising from dealing with specific properties of sequential data in general are presented in \cite{OberhauserKiraly} (recalled below for convenience). These can be addressed in a unified manner by using signatures \cite{FrizHairer,L14,LTL07}, which is the method of choice for generative modelling that we advocate in this paper. In Section \ref{sec:Signaturesfinancial} below we collect some compelling reasons to do so and also argue why a signature-based approach is tailor-made for financial applications such as pricing and hedging. We also indicate how a signatures-based objective function can be translated hedging performance of portfolios (Section \ref{sec:MMDIntro}).
\bigskip\\
\textbf{Problem formulation.}
Let us first fix some notations that will be used throughout the paper. In the following, let $S(t)$ denote the price of a financial asset (stock, exchange rate or index) and let  $X (t ) = \ln S (t )$ the corresponding log price.  Then the log return (at scale $t$) is denoted as
\begin{align}\label{eq:return}
r(t, \Delta_t):=X(t+\Delta_t)-X(t),
\end{align}
where the time scale  $\Delta _t$ (which we specify in Section \ref{sec:Methodology} for our methodology and experiments) has a scale ranging from a day to a month\footnote{In fact, $\Delta_t$ can range from a few seconds to a month, generally including certain high frequency applications as well. From a mathematical perspective there is no reason for us to exclude shorter scales, but in this analysis we focus on scarce data situations, hence we restrict one day to be the smallest unit.} Autocorrelation for a time lag $\tau>0$ is denoted by 
\begin{align}\label{eq:Autocorrelation}
\textrm{corr}\big(\ r(t +\tau, \Delta_t) \ ,\ r(t, \Delta_t) \ \big).
\end{align}
A numerical generation with an increased emphasis on the accuracy of the data generating process then aims at generating synthetically $M \in \mathbb{N}$ returns-sequences of length $k$ for a suitable $k \in \mathbb{N}$
\begin{align}\begin{split}\label{eq:sequencesgen}
(\widetilde{r_{1\ } }(t_1,\Delta_t)\ldots \widetilde{r_{1 \ }}(t_k,\Delta_t)), \ldots,
%\\ 
%\phantom{\widetilde{r_{1\ } }(t_1,\Delta_t)}\ldots %\phantom{\widetilde{r_{1 \ }}(t_k,\Delta_t)}\\
(\widetilde{r_M}(t_1,\Delta_t)\ldots \widetilde{r_M}(t_k,\Delta_t)),
\end{split}\end{align}
such that the generated set of $k$-sequences $(\widetilde{r_{i\ } }(t_1,\Delta_t)\ldots \widetilde{r_{i \ }}(t_k,\Delta_t))$ of returns reflects the properties of the observed $k$-sequences of returns 
\begin{align}\begin{split}\label{eq:sequencesobs}
(r_{1\ } (t_1,\Delta_t)\ldots r_{1 \ }(t_k,\Delta_t)), \ldots,
%\\
%\phantom{\widetilde{r_{1\ } }(t_1,\Delta_t)}\ldots 
%\phantom{\widetilde{r_{1 \ }}(t_k,\Delta_t)}\\
(r_N(t_1,\Delta_t)\ldots r_N(t_k,\Delta_t)),
\end{split}\end{align}
as \emph{accurately} as possible, where $N\in \mathbb{N}$ denotes the number of observations in the original dataset. \\
%In Sections \ref{sec:MMDIntro} and \ref{sec:MMDSignatures} we present an evaluation metric that is capable of fully determining the underlying distribution of the path generating process. As stated in \cite{OberhauserKiraly} targeting stylized facts is often ad hoc and therefore we suggest a more universal approach based on Signatures.\\
\smallskip\\
\textbf{Small data environments:} Note that typically, the number of generated samples $M \neq N$ need not be identical to the number of original samples $N$. A small data environment (see scenario (ii) in Section \ref{sec:PracticalApp}) would correspond to the situation where $M >> N$, with $N$ relatively small from a statistical or from a standard deep learning perspective\footnote{
Similarly, application (i) in Section \ref{sec:PracticalApp} could correspond to environments where $M \approx N$ are approximately the same. This kind of situation would also arise when simulating tick data for high frequency trading. Also Scenarios (iii) and (iv) of Section \ref{sec:PracticalApp} can occur with $M \approx N$ or $M >> N$.}. Such an example is the daily stock data or of leading indices (S\&P500, DAX, FTSE) where the number of data samples (dataset covers $~5000$ days worth of daily data) available for training is of orders of magnitude smaller than the amount of data normally needed in most neural network applications. In such situations the challenge is to efficiently extract the most relevant information from a small amount of available samples in a very simple generative network.
\smallskip\\ 
%Note also that a crucial part of the generative process is the choice of $k\in \mathbb{N}$ the length of the modelled sequence (see Sections \ref{sec:Methodology} and \ref{sec:LengthSeq}) as well as the form in which we encode of the sequences $(r_{i\ } (t_1,\Delta_t)\ldots r_{i \ }(t_k,\Delta_t))$ to train our generative model (see Sections \ref{sec:Methodology} and \ref{sec:RetvsSig}).\\
\noindent Clearly, the term above referring to the \emph{accuracy} of the modelling, is yet to be specified. In Section \ref{sec:Stylizedfacts} below, we recall a collection of properties that are widely accepted to be universal features (stylized facts) of time series of the form \eqref{eq:return} and \eqref{eq:sequencesobs}; and in Section \ref{sec:CommonEvMetrics} we briefly recall corresponding metrics that are commonly used to test for the presence of these styled facts. \\
\subsection{A reminder of specific stylised facts and evaluation metrics}\label{sec:Stylizedfacts}
Time series data of financial markets exhibits a set of stylised facts of financial markets, that a realistic financial market model is commonly expected to reflect. Below, we included a brief reminder to the most common ones, for more details see \cite{Cont00}.
\begin{enumerate}
\item[-] \textbf{Non-stationarity:} Financial time series are typically non-stationary, that is, past returns do not necessarily behave like future returns. The stationarity assumption states that for any set of time instants $t_1,\ldots,t_k$ and any time lag $\tau>0$ the joint distribution of returns is the same as the joint distribution of the lagged returns:
\begin{align}\label{eq:returns}
(r(t_1,\Delta_t)\ldots r(t_k,\Delta_t)) \ \sim \ (r(t_1+\tau,\Delta_t)\ldots,r(t_k+\tau,\Delta_t)).
\end{align}
This property is not guaranteed to hold for the returns process in calendar time, see \cite{Cont00}.
\item[-] \textbf{Heavy tails and aggregational Gaussianity:} Asset returns have (power-law-like) heavier tails than normal distribution, and have a distribution that is more peaked than the normal distribution. However, as the time scale $\Delta_t$ increases, the distribution looks more and more Gaussian.
\item[-] \textbf{Absence of autocorrelations of asset returns, but slow decay of autocorrelation in absolute returns:}\ Asset returns are uncorrelated (except for very short intraday timescales) but not independent. The autocorrelation \eqref{eq:Autocorrelation} function of absolute returns $|r(t, \Delta_t)|$ decays slowly, as a function of the time lag $\tau$ following a power-law.
\item[-] \textbf{Volatility clustering and multifractal structure:} Phases of high/low activity tend to be followed by phases of high/low activity, see  also \cite{GJR14,LRBM19}.
\item[-]  \textbf{Leverage effect:} Asset returns exhibits a leverage effect i.e. a negative correlation between  the volatility of asset returns and the returns process.
\end{enumerate} 
As mentioned above, the sequential nature of the data is usually modelled by an ad-hoc selection of \emph{stylised facts} and measured by corresponding evaluation metrics (see Section \ref{sec:CommonEvMetrics}) in the case of classical models. Due to the the lack of an established consensus for similarity metrics for sample paths,  to date this is also the case for generative models: 
The most straightforward optimisation routine is to target a number of essential stylized facts of the data in the training and in the evaluation of the generative procedure.
Handcrafted combinations of stylised facts and corresponding performance evaluation metrics can be included in the optimisation routine used in a specific structure, alongside with further objectives with regards to a specific application. 
Therefore, in such cases the chosen combination of optimisation objectives is often very specific to the application at hand and does not transfer easily to other applications. This is referred to in \cite{OberhauserKiraly} as the non-universality of the approach for financial time series. 
%With these preparations in mind, we can make the first conclusions that generative modelling for realistic financial time-series data is a challenging task that calls for bespoke methods.  

\subsubsection{Performance evaluation metrics}\label{sec:CommonEvMetrics} 
To date, the most commonly used evaluation scores include (but are not limited to) the following:\\
\begin{enumerate}
\item \textbf{Distributional metrics:} Such metrics target the cumulative distribution functions, in order to ensure that the distributions of the generated samples closely match the historical ones, often visualised by QQ-plots. To assess the goodness of fit, the difference between the historical sample distribution versus the distribution of the generated samples is measured with respect to a suitable metric $\mathcal{D}$ 
\begin{align}\label{eq:Distrmetric1}
\mathcal{D}_{b\in B} \ \big|F_{1,n}(b)-F_{2,m}(b)\big|,
\end{align}
where $n, m$ denote the number of original and generated samples respectively, and $B$ is (a suitable discretisation of) the sample space. 
%Note that if $\mathcal{D}$ is chosen to be the $\sup$-norm, the metric \eqref{eq:Distrmetric1} corresponds to the test statistic of the Kolmogorov-Smirnov test, and if $\mathcal{D}_{b\in B}:=\sum_{b\in B}$, then the metric \eqref{eq:Distrmetric1} resembles the distributional metric in \cite{WiBaiWoBu19} for an appropriate discretisation $B$ of the state space into bins.
\item \textbf{Tail behaviour scores:} Targeting properties of the underlying distribution that control the tail behaviour, higher order moments such as skewness and kurtosis. 
\begin{align}\begin{split}
\frac{1}{N_x}\sum_{j=1}^{N_x} \big| skew(x_{j})-skew([\hat{x}_{j}^{(1)},\ldots,\hat{x}_{j}^{(m)}]) \big|; \quad
\frac{1}{N_x}\sum_{j=1}^{N_x} \big| kurt([x_{j})-kurt(\hat{x}_{j}^{(1)},\ldots,x_{j}^{(m)}]) \big|
\end{split}
\end{align}
\item \textbf{Correlation-, and cross-dependence scores:} To detect serial autocorrelation in the time-series, and analogously for multidimensional time-series, cross-correlation scores. 
%\begin{align}
%.
%\end{align}
\end{enumerate}
The above examples already suggest a wealth of possible metrics and evaluation scores to measure the quality of generated market paths and the list does not end here, depending on the application one has in mind:
While the above evaluation scores check effectively for some of the most relevant stylized facts, we might be interested in other properties of the generated data as well: We may want to optimize for example with respect to the expected payoffs of vanilla options, or of a portfolio of options.  
Or we may want to optimize the generative process with respect to the P\&L of the hedged portfolio under an appropriate risk measure. Having deep hedging in mind as an application for the generative model, we may aim to include hedging objectives in the optimisation either directly (which may be computationally expensive) or indirectly.
For the latter we refer to \cite{AntonovQuantifying18} for some possibilities and we refer to Section \ref{sec:MMDIntro} for a brief explanation on how our suggested performance evaluation metric (see (\emph{Step 5}) of Section \ref{sec:MainResults}) links back to hedging strategies.

\subsection{Challenges for similarity metrics for financial time-series by generative models}\label{sec:ChallengesSimbyGen}
Traditional distributional metrics and divergences provide ample means to measure distances between distributions. But determining appropriate similarity metrics on the level of a stochastic process---or its data representation through a time series---is a challenge of a different nature. This section is devoted to the questions: What are potential challenges in determining appropriate metrics to measure the similarity of distributions on path space, or to determine whether two sets of sample paths originate from the same underlying distribution?
Generalisations of the static case (similarity of marginal distributions) to the dynamic one (similarity metrics on path space) are not straightforward. 
Similarly,  evaluating the ``goodness'' of a generative model for sample paths of financial time-series is a challenge in particular. With view to our goal of proposing effective performance evaluation metrics for market generators we list a number of these challenges that are addressed in this paper:
\bigskip\\
\textbf{(1)} The potential non-universality (as described in Section \ref{sec:CommonEvMetrics} above) of features to control for in the generated time-series. If the chosen set of optimisation objectives is bespoke to one application, the generated time series may not carry over easily to other applications.\\
\textbf{(2)} The underlying distribution of the data generating model is often not known explicitly in generative models (See Section \ref{sec:Introduction} above) and only controlled implicitly through the generated data samples and their similarity to the original data. 
Therefore, \emph{two-sample tests} may be better suited as performance evaluation metrics than distributional metrics and divergences (KL-divergence, Fisher Information metric, Wasserstein metric). The latter may only be applicable after inferring from the available data sample to the underlying distribution.\\
\textbf{(3)} Established distributional metrics (and two-sample tests) for marginal distributions can be generalised to (finite) multivariate marginals. However, generalising these metrics to path space is not straightforward, one of the difficulties being that the  pathspace $C^0([0,1], \mathbb{R}^d)$ is infinite dimensional and non-locally compact, see \cite{CO18}.\\
\textbf{(4)} Non-continuous observation of the original data.  Usually only discrete time observations of sample paths are available. While for classical models this is less problematic, for generative models an appropriate feature set needs to be specified. Modelling sample paths as a learning problem of vector-valued data is problematic for several reasons: the number $n$ and position of observation time points might change from sample to sample. Also, in some applications (for example high-frequency data) $n$ can get very large. Finally, from a hedging perspective it is delicate (and often not sufficient) to match marginal distributions on a finite number of observation dates only. As shown in \cite{Brigo19} one can construct examples that are indistinguishable from one another on a finite set of marginals in the physical measure, but lead to arbitrarily different hedging strategies and option prices.\\
\textbf{(5)} Non-stationarity of the target distribution. Financial time-series are typically non-stationary and underlying distributions change with market conditions. For generative models this has two implications in particular: It is beneficial to design a \emph{conditional} version of the market generator, which allows to produce samples that are conditioned on specific market states, see Section \ref{sec:MainResults} for more details. Furthermore, training a conditional generative model may amplify the data scarcity issue as the availability of sufficient number of representative training samples becomes coupled with market conditions.
\subsubsection{Two-sample tests and Maximum Mean Discrepancy metrics as performance evaluation metrics}\label{sec:MMDIntro} 
If the distributions of the original data generating process $({r_{i\ }}(t_1,\Delta_t)\ldots {r_{i \ }}(t_k,\Delta_t))$  and the generated series $(\widetilde{r_{i\ } }(t_1,\Delta_t)\ldots \widetilde{r_{i \ }}(t_k,\Delta_t))$ are known explicitly, a number of established similarity metrics can be computed efficiently. In practice however, this underlying distribution of the original data is (often) only available through the given samples and not known explicitly. 
For generative models, the same holds for synthetic samples as well as summarised in \textbf{(2)} above. If at least one of these data samples is small, metrics on the level samples rather than on the level of distributions are preferable.\\ \emph{Two-sample tests} provide a flexible framework to compare the empirical distributions of two given data samples according to the following principle:
If $X$ and $Y$ are random variables with respective probability measures $p$ and $q$ defined on some common state space; given i.i.d. observations $\{x_1\ldots,x_m\}$ and $\{y_1,\ldots,y_n\}$ from $p$ and $q$ respectively, can we decide whether $p\neq q$?\\ Typically these tests are designed to determine whether two given data samples were generated by a common underlying distribution or not, but do not specify what that common distribution is\footnote{
One of the best known such nonparametric two-sample tests is the popular two-sample Kolmogorov-Smirnov test, but it has some shortcomings: It may need a large number of samples and it is difficult to generalise to higher dimensions.
%More powerful two sample tests are the Maximum Mean Discrepancy (MMD) tests, which also have advantages in terms of better interpretability and stability.
}.
A number of related tests addressing matter \textbf{(2)}, have recently become more popular in generative modelling, see \cite{MMD1}: In \cite{MMD1}, the authors determine if two samples are drawn from different distributions $p$ and $q$ based on the largest difference in expectations over predefined set of functions in the unit ball of a Reproducing Kernel Hilbert Space (RKHS) $\mathcal{H}$.
\begin{align*}
\textrm{MMD}[\mathcal{H},p,q]:=\sup_{f\in \mathcal{H}}\left(\mathbb{E}_{X\sim p}\left[f(X)\right]-\mathbb{E}_{Y\sim q}\left[f(Y)\right]\right).
\end{align*}
Where the space $\mathcal{H}$ is rich enough to distinguish between metrics $p$ and $q$. Therefore the metric is called the maximum mean discrepancy (MMD). Though this does not yet address the matter \textbf{(1)} of non-universality, using these metrics in the training phase have rendered more stable convergence results in several studies and they can be computed efficiently.
Therefore, MMD quickly entered the neural network arena\footnote{In \cite{MMDWasserstein} MMD has been established in the context of Wasserstein Auto-Encoders and \cite{MMDGan} introduce MMD-based GANs. By making fundamental connections to optimal transport distances in the data space, MMD measuresestablish the theory proving the correctness of this generative procedure. } in generative modelling. \\
This is the performance evaluation metric we use for our generative algorithm, see  \eqref{eq:MMDSig} in Section \ref{sec:MMDSignatures}. Generalising Maximum Mean Discrepancy metrics to path space faces the same challenges as the ones summarised in \textbf{(3)}. In  a recent paper \cite{CO18}, Chevyrev and Oberhauser develop a Maximum Mean Discrepancy based two-sample that relies on a feature map from stochastic analysis called the ``\emph{signature}'' of a path, which tackles point \textbf{(3)} simultaneously with the issue \textbf{(1)} of non-universality. In fact, this test is based on a notion that is reminiscent of the role of moment generating functions on path space and hence characterises the distribution of the stochastic process uniquely, see Appendix \ref{sec:Signatures} for details. 
Furthermore, it follows from \cite{ABBC18,L98}, that in addition to the advantages above, signatures also provide the right framework to match hedging objectives and hence to bypass matter \textbf{(4)}.
\smallskip \\ In fact, the insight that the similarity on the level of signatures (hence passing the signature based MMD test described in Section \ref{sec:MMDSignatures}) can be linked back to similarity in hedging performance is a consequence of the continuity propertiy of the It\^o-Lyons map \cite{L98}. This states that the mapping from the driving path controlled differential equation to its solution is continuous (in fact Lipschitz) in a suitable rough path norm. As a consequence, if the signatures of two driving paths are \emph{close}, then so are the solutions of the corresponding controlled differential equations:
Since the performance of a hedging strategy can be written in form of a controlled differential equation (the performance is an It\^o integral of the hedging strategy against the price process) then if two price paths have similar signatures (i.e. pass the  signature based MMD test described in Section \ref{sec:MMDSignatures}), it follows that they will have similar performance under the same hedging strategy.

%********************\\
%A few important points specific to path-valued data are (cf. \cite{CO18})
%\begin{itemize}
%\item for many real-world signals, it is beneficial to ignore parameterization of time; e.g. in speech, different speakers pronounce the same words at different speeds.
%%\item usually only discrete time observations $(x(t_i))_{i=1,...,n}$ are provided (due to finite storage, sampling cost, etc). However, treating this as a learning problem of vector-valued data is problematic since the number n and position of time points might change from sample to sample. Additionally, n can get very large for high-frequency data.
%\item unbounded variation paths arise by Donsker-type theorems in the high-frequency limit and functions of such paths have to be treated with care. For example, important functions such as quadratic variation, or the solution of nonlinear filtering or a stochastic differential equation, do not depend continuously on the underlying path. (cf. \cite{CO18})
%\end{itemize}
%The framwork provided by signatures is parameterization invariant (with the option to capture parameterization variance); it is robust to irregular sampling a path at irregular times; it appears naturally when describing the behaviour of functions of non-smooth paths.\\
%********************\\
%********************

\subsection{On signatures, their advantages in encoding path valued data, and their meaning}\label{sec:Signaturesfinancial}
The challenges that generative modelling of financial data streams faces---raised in the beginning of Section \ref{sec:challenges}---are not limited to the choice of appropriate performance evaluation metrics only. They also extend to training: An efficient parsimonious encoding (feature map) of financial data streams also enhances the training of such generative models. Furthermore, an encoding that anticipates typical properties of the data and irregularities of sampling, makes training more robust with respect to data quality.  The framework of rough paths and signatures \cite{FrizHairer,GLKF14,L14,LTL07,OberhauserKiraly} lends itself well to these aims as well: They provide a means to encode financial data streams parsimoniously and efficiently and they a powerful framework to address further challenges of path valued data as well. Therefore, we do not only use the signature framework for performance evaluation but also resort to these as a feature map in the generative model itself. 
\begin{definition}[Signature of a path]
Let $X:[0, T]\to \mathbb R^d$ be a continuous path of bounded variation. Then the signature of $X$ is defined by the sequence of iterated integrals given by
\begin{align}\label{eq:signature}
\mathbb X_{T}^{<\infty} := (1, \mathbb X_{t}^1, \ldots, \mathbb X_{T}^n, \ldots)
\end{align}
where
\begin{align}\label{eq:signatureentries}
\mathbb X^n_{T} := \int_{0<u_1<\ldots<u_k<T} dX_{u_1}\otimes \ldots \otimes dX_{u_k}\in (\mathbb R^d)^{\otimes n}
\end{align}
where $\otimes$ denotes the tensor product. Similarly, given $N\in \mathbb N$, the truncated signature of order $N$ is defined by
\begin{align}\label{eq:truncsignature}
\mathbb X_{T}^{\leq N} := (1, \mathbb X_{T}^1, \ldots, \mathbb X_{T}^N).
\end{align}
\end{definition}

\begin{remark}
If the path $X$ has bounded variation -- which is the case of discrete data -- the integrals above can be defined using Riemann-Stieltjes integrals.
\end{remark}

\noindent %Classical generative models in machine learning  typically provide tools to learn probability distributions on finite-dimensional spaces. 
%In the context of generative models for financial data however, the probability distributions in question is defined on an infinite-dimensional space i.e. the path space $C^0([0,1], \mathbb{R}^d)$. \\
As mentioned in the beginning of Section \ref{sec:challenges}, a particular challenge in the context of synthetic generation of market paths, is that the distribution in question is defined on the infinite-dimensional space of paths $C^0([0,1], \mathbb{R}^d)$, while the available generative modelling tools are finite-dimensional. Thus the infinite-dimensionality of path space is not only a challenge for the choice of suitable similarity metrics or performance evaluation metrics but also for feature extraction of the data and training. A solution to this issue is to project this infinite-dimensional space to a suitable finite-dimensional space where standard methods for generative models may be used.\\
However, mapping this inherently infinite dimensional space in an optimal way to a finite dimensional one presents a challenge and the choice of projection is not trivial:\\
The most straightforward  choice would be to sample the path on a fixed, discrete time grid return by return  as in \eqref{eq:return} for example (while accounting for the joint distribution of these), and learn the projected probability measure using standard generative models. This approach would not fully capture the sequential nature of financial data and would fail to effectively capture the probability measure on the original path space. Moreover, if we project this infinite-dimensional object down to a finite-dimensional space by sampling on a discrete time grid, the projection is not a ``natural one'' as the original distribution on an intrinsically infinite-dimensional space captures much richer information about the process. The latter can have significant consequences on hedging as outlined in \cite{Brigo19} and in the end of Section \ref{sec:MMDIntro} above. \\ 
A more effective projection is to use the \textit{signature} or \textit{log-signature}\footnote{See Definition \ref{def:logsig} in Appendix \ref{sec:Signatures} for the latter.} to project an infinite dimensional encoding \eqref{eq:signature} of the path space to a finite $N$-dimensional one as in \eqref{eq:truncsignature}.
The signature of a path is a transformation of the original continuous path into a sequence of statistics, an infinite dimensional vector \eqref{eq:signature} of signature entries \eqref{eq:signatureentries}. These statistics fully characterise the original path up to time parametrisation (see \cite{HL10,horatio} and  Appendix \ref{sec:Signatures} for details) and furthermore they offer a faithful and parsimonious description of it already in its first few entries (dimensions) of the signature vector \eqref{eq:truncsignature}. The error made by the truncation at level $N$ decays with factorial speed as $\mathcal{O}(1/N!)$, see \cite{LTL07}.
\bigskip\\
Moreover, the first several signature entries---i.e. the first terms in the vectors \eqref{eq:signatureentries} resp. \eqref{eq:truncsignature}---have clear financial interpretations:
The first term captures drift -- i.e. the increment of a price path over a period of time. 
The second term indicates the volatility over the period of time (through the L\`{e}vy area). Higher order terms capture finer aspects of the path that end up fully characterising the latter. An ordering, reminiscent of principal components from the most relevant towards finer properties of the path. See the Appendix \ref{sec:Signatures} for more details.
\subsubsection{On returns-based versus signature-based  data generation in a pricing and hedging context}\label{sec:retsig} It clear from the previous section that the signature transform is a highly efficient way of encoding the most relevant information contained in a stochastic path. We demonstrate in our numerical results below that learning the (truncated) signature of a set of paths leads to a more efficient learning (i.e. training converges with fewer training samples already) than learning the multidimensional distribution of the process on a discrete grid. The projection on the finite dimensional space of truncated signatures is not only numerically more efficient than the finite dimensional projection on a discrete time-grid. The signature-based projection encodes a richer and more relevant wealth of information about the path (which also allows us to control for option prices and hedging strategies \cite{ABBC18,IPA18}), while in the latter projection some essential information may be lost, which can have financial consequences on option prices and hedging strategies: In fact \cite{Brigo19} shows that when sampling returns distributions of a stochastic process on a discrete time grid, even statistically indistinguishable sets of paths in the historical measure can lead to arbitrarily different option prices. If however the paths are sampled on the level of signatures, this ambiguity of option prices does not occur. The findings of \cite{ABBC18,Brigo19} therefore indicate that not only is the signature a more efficient way of encoding sample paths 
but one that that removes the ambiguity of the corresponding option prices and provides a meaningful control over hedging performance. See the end of Section \ref{sec:MMDIntro} above for the last statement.
\subsubsection{Further advantages of signatures} Further advantages of working with signatures for modelling functions of data streams have been discussed and presented in \cite{L14,OberhauserKiraly}. These advantages include (but are not limited to) the following properties:
The expected signature of a stochastic process \textbf{determines the law of the process uniquely}. With that, the expected signature plays a similar role on path space as the moment generating function for distributions
(this provides a basis of the performance evaluation metric developed in \cite{CO18} for path space).\\
\textbf{They permit a model-free data-driven modelling:} The framework does not impose any assumptions on the underlying stochstic dynamics. Signatures provide a flexible basis of functions for a functional on path space. But while Fourier transforms and wavelets have a similar role approximating curves as a linear combination of basis functions, signatures do so in a model free, unparametrised way (since a path by path characterisation is possible).\\
\textbf{The signature transform is traightforward to implement:} Today there are readily available (and constantly improving) powerful python packages and libraries\footnote{Such as {\tt esig, tosig} and {\tt iisignature} libraries, see \href{https://github.com/bottler/iisignature}{https://github.com/bottler/iisignature} as well as \cite{KiLy20} for more background. }  to transform data streams to signatures and algorithms to transform signatures back to paths of datastreams. For the inverse transform, one possible algorithm is developed in this paper (see Section \ref{sec:Methodology}) and  provided in the Github repository \href{https://github.com/imanolperez/market_simulator}{Github:Marketsimulator}.\\
The framwork is \textbf{invariant under translation\footnote{Signatures  are constructed from increments of the path. As such, they are also invariant under translation: all reference to absolute values of the path is lost. One may find settings where it is important to make reference to absolute values of the path; e.g. to ensure that the asset price remain positive. In such cases, the so-called \emph{visibility transform} can be used.} and time-parametrisation
.}
Therefore, in order to encode price paths in business time rather than calendar time we apply the lead-lag transformation\footnote{
One of the commonly used method for measuring similarity between two temporal sequences, which may vary in speed is \emph{Dynamic time warping (DTW). A well known application has been automatic speech recognition, to cope with different speaking speeds. }
In finance a similar consideration suggests measuring business time rather than calendar time of the process.} (see Section \ref{sec:leadlag} for more details).\\
Furthermore, signatures are \textbf{robust to irregular sampling} (which becomes relevant for tick-data),  \textbf{missing data and towards highly oscillatory data as well}: In particular, they provide a consistent framework for unbounded variation paths, which may arise by Donsker-type theorems in the high-frequency limit. Functions of such paths have to be treated with care, as for example the quadratic variation, or the solution of nonlinear filtering or of  stochastic differential equations, do not depend continuously on the underlying path. Signatures appear naturally when describing the behaviour of functions of non-smooth paths, cf. \cite{CO18}. 
\subsubsection{From signatures to log-signatures}\label{sec:SigtoLogsig} In the present work, our generative model we are not targeting the signature directly, instead we first use a bijection to log-signatures (See Definition \ref{def:logsig} in Appendix \ref{sec:Signatures} for the latter). Generative modelling on signature space directly may be problematic, which results from the fact the \textit{signature space} is not linear: Therefore, small perturbations of the signature (as one might expect to obtain from the output of a generative model on signature space) of a path  will in general not correspond to the signature of some other path. In fact there may not exist any path with a signature that results from the perturbation.  We solve this issue by working with the so-called \textit{log-signature} instead, which also characterises the price path, but now spans a linear space.  We refer the reader to \cite[Section 1.3.5]{primer} for a detailed discussion of the log-signature, as a full reminder of its formal characterisation and properties is out of the scope of this paper.
For more information on the background and implementation of signatures and log-signatures, available python packages and libraries for signatures, log-signatures and the lead-lag transformation see the related works \cite{primer, KiLy20, NiLy19, ReizGraham18} such as a brief reminder in the Appendix \ref{sec:Signatures} below. Recent machine learning applications using signature inputs include \cite{BKPSL19,CKT20,GKMO18,L14,NiLy19,OberhauserKiraly}.

\section{Main results: Our methodology and its background and motivation}\label{sec:MainResults}
\textbf{Problem setting and main results and contributions:} Given a financial data stream\footnote{We assume the data stream to be univariate for notational simplicity but it is straightforward to extend the methodology can be extended to multivariate data streams too.}  (be it a market index S\&P, DAX or a numerically generated series) we work with the one available realisation of the evolution of this stream. From the observed path observed on a time horizon of several years we intend to infer the underlying data generating process in order to produce further samples from that distribution\footnote{For this we temporarily make the customary stationarity assumptions which we will later relax.}.
We train a variational autoencoder (VAE) to reproduce the underlying distribution of an observed spot index on different time horizons up to a month. Note that our methodology is by no means limited to one-month time horizons and in the postprocessing step (Step 4) of our methodology we propose a way to generate data  on longer time horizons (here up to several years) by appropriately concatenating paths.
\begin{enumerate}
\item[(a)]As a first step we demonstrate that a returns generation in the classical sense is possible by variational autoencoders on various time-scales (daily, weekly, monthly returns). This experiment follows the spirit of other currently available market generation approaches, and supports our choice of VAE as a parsimonious generative model. 
\end{enumerate}
We then also go beyond simple returns generation in the following ways:
\begin{enumerate}
\item[(b)]\label{pathspace} To accommodate to inherent infinite-dimensional nature of the problem, we propose a generative process directly on path space, based on the signatures (see Section \ref{sec:Signaturesfinancial} above) of the paths. 
%This has the advantage 
We then demonstrate that the generated paths do not only fit all the observed marginal returns distributions (obtained in the approach above) but also the joint distributions of returns on these time-scales and other essential features of the data.
\item[(c)]\label{conditions}  We fine-tune the generative process by allowing conditioning on various market indicators to account for the non-stationarity of the time series. With this conditioning we refine our (returnes-based or signature-based) variational autoencoder (VAE) to a conditional variational autoencoder (CVAE), and  generate paths conditional on various market states. The latter conditioning also allows us to paste path segments conditioned on the signature of the previous path segment, and thereby generating paths of arbitrary length.\\
\end{enumerate}
\newpage
\subsection{Methodology: An overview of the main steps}\label{sec:Methodology} 
Our algorithm and numerical experiments can be subdivided into the following five main steps. We first give a brief overview of these and provide in Section \ref{sec:BackgMotMethodology} more explanations on each of the steps.\\
\begin{enumerate}
\item[(Step 1)] \textbf{Data extraction from time series: } 
In our experiments given a sample path from a data stream, we subdivide the the full time series in equal length intervals\footnote{For the numerically generated data in our experiments, we directly generate sample paths of the above length (i), (ii) and (iii), but we could have applied the same method as above.}. Here we subdivide into segments of: (i) 1 day, (ii) 5 days, corresponding to a business week, and (iii) 20 days corresponding to a month.
\item[(Step 2)]  \textbf{Preprocessing the data:} To obtain training data from the resulting path segments, we 
\begin{itemize}
\item[(a)] calculate log-returns $r(t, \Delta_t)=X(t+\Delta_t)-X(t)$ with appropriate $\Delta_t$, %cf. \eqref{eq:return} 
to generate \\(i/a) daily log-returns, that is $\Delta_t=$ 1 day,\\ (ii/a)  weekly log returns, that is $\Delta_t=$ 5 days and \\
(iii/a) monthly log returns, that is $\Delta_t=$ 20 days, from the data. 
\item[(b)]convert the obtained data samples into log signatures (for the paths of length (b/ii) 5 days and (b/iii) 20 days) applying the lead-lag transformation. This will enable the  pathwise generation process described in point (b) above. A mathematical background and motivation for these transformations is given in Appendix \ref{sec:Signatures}.\\
\end{itemize}
\item[(Step 3)] \textbf{Creating and training the VAE network:} After splitting the historical data into training/testing/and validation sets 
\begin{itemize}
\item[(a)] we train a variational autoencoder \textbf{VAE(a)} on the daily, weekly and monthly returns for (i/a), (ii/a), (iii/a). The output of this VAE is on the level of returns.
\item[(b)] We train a variational autoencoder \textbf{VAE(b)} on the log-signatures of weekly and monthly sample paths for (ii/b), (iii/b) (see Sections \ref{sec:Signatures} and \ref{sec:signatureproperties}). The output of this generative VAE(b) is then given in form of log-signatures.
\item[(c)]In the refined version we also calculate and store relevant market conditions such as current level\footnote{That is, the instantaneous level at the start each path segment.} of volatility, current level of the index, signature of the previous path segment. These values will then be used in the CVAE (Conditional Variational Autoencoder) to generate new data points conditional on these indicators.\\
\end{itemize}
\item[(Step 4)] \textbf{Postprocessing of the outputs of the VAEs:} At this stage of the generative process we either convert back the generated log signatures into paths or use the generated (log-)~signatures directly. In fact, in case the purpose of the generative process to provide training data for neural networks with pricing and hedging objectives, both options are available: Either: (4.b) Signatures can be used for option pricing directly as suggested in \cite{IPA18}, Or: (4.a) In order to invert signatures into paths, we suggest in Appendix \ref{sec:Signatures} one possible method do so\footnote{Note that developing the computationally most efficient ways of this task are mathematically highly nontrivial, and subject of ongoing research \cite{C18}.}. In fact, inverting signatures into paths is also instrumental to allow comparing the performance of the generative networks on the level of returns distributions, as described in the second table of the following point.\\
\item[(Step 5)]  \textbf{Performance evaluation:} We evaluate the performance of each of the above approaches to the generative process and compare the outputs (both to the original data samples and to one-another) with respect to different similarity metrics and conclude that the signature-based generation outperforms the returns-based approach. In fact in case the purpose of time-series generation lies in the context of pricing and hedging derivatives, the suitable similarity metrics are indeed the signature-based similarity metrics (see Sections \ref{sec:MMDSignatures} and \ref{sec:MMDIntro} for details). 
%while returns-based similarity metrics cannot rule out arbitrarily large differences in pricing even if a perfect match is achieved on the level of marginals.
This follows directly from findings of \cite{Brigo19}.\\ \\ 

\subsection{Background and explanations to our generative modelling methodology}\label{sec:BackgMotMethodology}

\subsubsection{(Step 1) Data extraction from time series stream}\label{sec:LengthSeq} It directly follows from the non-stationarity of financial time series that `past returns do not reflect future performance' \cite{Cont00}. Therefore, strictly speaking, the data of the evolution of an index over time amounts to one single observation of a realised path. However as a starting point to any perform statistical analysis of market data one needs several observations of the underlying quantity to which one of the most basic requirements is `the existence of some statistical properties of the data under study which remain stable over time' \cite{Cont00}. To extract multiple observations from this data stream, one divides this path into segments. This (as usual) calls for some sort of stationarity assumption, though financial time-series are known to be typically non-stationary.
In our experiments given a sample path from a data stream, we subdivide the the full time series in equal length intervals: (i) 1 day, (ii) 5 days,  and (iii) 20 days.
In dividing the observed path into segments, there are different considerations to balance with one-another:
\begin{itemize}
\item The longer the path segments, the fewer  path segments are available\footnote{
A 20 year daily observation timeline results in $\sim 250$ sample paths with monthly (20 days) path segments, or $\sim 1000$ weekly (5 days) path segments or in $\sim 5000$ daily returns.}.
\item The longer the path segments the less severe the violation non-stationarity assumption in the obtained training data. 
\end{itemize}

\subsubsection{(Step 2) Returns-based (a) versus signature-based (b) data generation:}\label{sec:RetvsSig} 
 If one wants to create a generative model that learns to generate samples from a distribution on path space, one has to decide what representation of the path will be used: a returns-based (a) or signature-based (b) projection from path space to a finite dimensional space. This representation has to be an effective representation of the path, and it should be rich enough to capture the distribution of the paths. 
See Section \ref{sec:Signaturesfinancial} for a discussion of compelling reasons to use the truncated signature projection for this purpose. Given that the signature of a path uniquely determines the path \cite{BoedGe15} and that the expected signature uniquely determines the distribution of the paths \cite{CL16}, it is natural to use the signature to represent the path. When the truncated signature is used, the representation of the path is essentially a vector on some high-dimensional space. This transformation of paths to signatures can then be applied to our sample of paths, in order to then use traditional generative models such as Variational Autoencoders. However, the signature of a path is a group-like element \cite[Definition 2.18]{LTL07} whereas the generated signatures won’t. Therefore, it is more convenient to apply the generative model to log-signatures, because once we compute the tensor-algebra exponential of the generated log-signatures the resulting element will be group-like \cite[Theorem 2.23]{LTL07}. In this case, the output of the generative model will be in form of log signatures as well and may have to be inverted to paths in a post-processing step (see \emph{Step 4}).
In fact, the generation of paths in (log-)signature space is more efficient than maximixing a corresponding fit in returns for a simulated stochastic process in the sense that the generative process achieves a higher precision already with less training data. This can be seen in comparative performance metrics \emph{(Step 5)} in the numerical experiments:
There, a comparison with purely returns optimised VAE shows the proof of concept that signatures work better. See Section \ref{sec:pathsrealisticMMD} for the corresponding numerical results.
More precisely, the comparison of the weekly 5-dimensional joint distribution with the weekly signature paths we show that the signature-based generation outperform the weekly joint distribution-of-returns based gneration. The reason for this is that signatures are an efficient feature map to encode the most relevant information captured in time series data.\\

In a pricing and hedging context there is an even more compelling reason to use signatures as a feature map for the time series: Not only does a signature-based generation yield a faster training and more stable convergence, but in a pricing and hedging context it eliminates pricing ambiguities while pure returns-based generation does not as Brigo explains in \cite{Brigo19} in a framework consistent with statistical analysis of historical volatility that can lead to arbitrarily different options prices.
Previously, \cite{ABBC18, Brigo19} had shown that implied volatility is linked with a purely pathwise lift of the stock dynamics, confirming the idea that while historical volatility is a statistical quantity, implied volatility is a pathwise one. See also the end of Section \ref{sec:MMDIntro} for a hedging perspective.

\subsubsection{(Step 3) Network Architecture and Training of the VAE and CVAE}
The Achitecture of the VAE and CVAE networks such as the market indicators that serve as conditioning variables for the CVAE are summarized in this section below. For full details of the model and training we refer the reader to the Github repository \href{https://github.com/imanolperez/market_simulator}{Github:Marketsimulator}.\smallskip\\
\textbf{The encoder:} The encoder network has one hidden layer and two latent layers, with 50 nodes on the hidden layer. The activation function is a leaky (parametric) ReLU with parameter $\alpha=0.3$.\\
\textbf{The decoder:} The decoder network has ne hidden layer with 50 units and activation function leaky (parametric) ReLU with parameter $\alpha=0.3$. The output layer with sigmoid activation function.

\subsubsection*{Conditional Variational Autoencoder, adapting to specific market conditions}
In order to further accommodate to the non-stationarity of the data, we now refine the VAE to certain specific market conditions: a Conditional Variational Autoencoder (CVAE). The market conditions we consider here are the following:
\begin{enumerate}
\item \textbf{Level of the instantaeons volatility at the start of the path.}
\item \textbf{Level of the index at the beggining of the path.}
\item \textbf{Log-signature of the previous path:} This last condition is in fact more refined than the first condition, in fact it is more restrictive. If we control for the log-signature of the previous path, we automatically control for the volatility.
See Section \ref{sec:Signaturesfinancial} for more details on the financial interpretation of the elements of the signature vectors.
\end{enumerate}

\subsubsection*{Motivation for VAE as our generative model} 
Here, we briefly motivate our choice for VAEs over the other generative modelling approaches: VAEs aim at maximizing the lower bound of the log-likelihood of the observed data, see Appendix \ref{sec:VAE} for details. With that they are parsimonious, theoretically clear to explain and also easy to implement and to interpret.
In summary, VAEs are stable and flexible generators for scarce data environments. Furthermore, VAE frameworks work consistently well under  different differential operators and architectures including recurrent networks, \cite[Chapter 20]{MLBook}.
%, while Boltzmann machines require careful model design to maintain tractability. 
This flexibility may prove useful in later applications.\\
Arguably, the most popular differential generator networks today are GANs. In fact, the performance evaluation metric presented in (\emph{Step 5}) can be seen as a non-automated discriminator applied as a one-step verification that the generated samples are indistinguishable from the original ones. \\
Our choice for the VAE approach in the current context is based on the following considerations.
\begin{itemize}
\item The relative unpopularity of VAEs stems from image processing and its relative weakness in that context are irrelevant for time-series applications. In fact, the relative unpopularity of VAEs compared to GANs in image processing originated from the fact that generated samples from VAEs trained on images tend to be somewhat blurry. This could be attributed to the fact that the model (maximising the likelihood of the observed dataset) may attribute a high probability to (nearby) points other than the ones in the training set. This however would be no drawback of the VAE in a time-series setting. 
\item VAEs require considerably less data for training than GANs. A GAN aims at achieving an equilibrium between a \emph{Generator} and \emph{Discriminator}. Achieving the equilibrium between generator and discriminator networks of GANs is in most practical scenarios a highly delicate matter and often sensitive to hyper-parameter tuning (cf \cite[p. 692]{MLBook}).
%\footnote{Goodfellow (2014) identified nonconvergence as an issue as simultaneious gradient descent is not guaranteed to reach an equilibrium. It is not known to what extent this nonconvergence problem affects GANs. Furthermore, also stabilisation remains an open problem (cf \cite[p. 692]{MLBook}).}.  
Recently much research activity was devoted to addressing this issue, and reformulations of GANs have been proposed that are guaranteed to converge, if provided with enough training samples. 
%In spite of the recent research efforts and progress, GANs remain to be unstable, sensitive to hyperparameters and notoriously data-hungry. 
Since in this application we are specifically targeting scenarios where training data is scarce, we opt for generative variant of Autoencoders (the nonlinear generalisation of PCA, see \cite{MLBook}) for generative modelling in our current context.%\footnote{The underlying principle of a (vanilla) autoencoder is to encode an input image to a much smaller dimensional representation which can store latent information about the input data distribution. But in a vanilla autoencoder, the encoded vector can only be mapped to the corresponding input using a decoder. It certainly can’t be used to generate similar images with some variability.}.\\
\end{itemize}

\subsubsection{(Step 4) Postprocessing of the VAE outputs}\label{Sec:Postprocessing}
%\subsubsection*{Inverting back from Sigatures to paths: Inverting logsignatures} 
When the Market Generator is a VAE(b)-type of generator, that is the outputs of the generative process are given in form of log-signatures or conditional log-signatures (conditioned on the log-signature of the previous path segment), then:
\begin{enumerate}
\item \textbf{We can leave the output of the (C)VAE in the form (log-)signatures and apply it directly to pricing:} Signatures can be used directly for pricing vanilla products or exotic derivatives as proposed in \cite{IPA18} and this approach has several advantages in particular for computation heavy, path-dependent applications. Therefore, if we have these algorithms in place, for many applications (including XVA oriented computations) it is advantageous to leave the output of our generative model on the log-signature level and apply directly pricing algorithms for signatures to that.
\item \textbf{Alternatively, we can convert back the the output of the VAE(b) from signatures to sample paths:} In this paper we develop an evolutionary algorithm for this purpose. Other alternatives are currently a topic of active research \cite{C18}.
The idea behind inversion of signatures to paths is based on the following idea: The signature of a path uniquely determines the path itself \cite{HL10,horatio} -- in other words, if we know the signature of a path, we know the path itself. However, the task of retrieving the path from its signature (or log-signature)in a computationally efficient way is a highly non-trivial task \cite{C18}. In this paper, given that stock prices discrete (multiples of the pip size), we used an evolutionary algorithm to retrieve a path whose log-signature is close to the generated one.
Evolutionary algorithms aim to solve certain optimisation problems by mimicking to some extent biological evolution. In the context of signature inversion, we start with an initial population of random paths, and we iteratively (i) select the paths whose signatures are closest to the target signature, and (ii) \textit{breed} these paths and introduce \textit{mutations} to generate a new generation of paths. After sufficiently many iterations, we end up with a population of paths whose signature are close to the target signature we aim to invert.

\item \textbf{Concatenation of paths for longer time-horizons:} Recall that in VAE(b/ii) the outputs of the generative process are log-signatures of weekly paths and in VAE(b/iii) the outputs are log-signatures of monthly paths. Recall also, that the longer the paths, the less samples we have available. However, in some scenarios we might be interested in obtaining longer sample paths than the outputs of the generative model: Say, generating signatures of monthly (or even longer) paths from the outputs of VAE(b/ii) the weekly signature-based generative model. This can be done, using the multiplicative algebra structure of the signature space.
We demonstrate this by assembling monthly signature outputs VAE(b/iii) to yearly paths in our numerical experiments, see Figure 1 below. 
\begin{figure}\label{fig:Yearlypaths}
\includegraphics[scale=0.5]{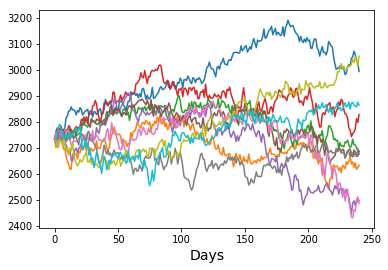}
\caption{Here, we demonstrate that the method can be applied to generate paths on longer horizons than one month (here up to one year) by concatenating generated paths with the CVAE where we consecutively condition the output of the monthly log-signature CVAE generation on the log signature of the previous segments (generated) path's signature. See \emph{item (c)} of  \emph{(Step 4)} in Section \ref{Sec:Postprocessing}. \vspace*{0.3cm}
}
\end{figure}\\
\end{enumerate}
\textbf{Converting from returns (on different time scales) to paths}: If the Market Generator is a VAE(a)-type of generator, that is, the output of the VAE is in form of returns, one may be interested in producing sample paths from these (i) daily,(or the (ii) weekly or (iii) monthly returns) generated returns. We do so in our numerical experiments in order to compare the performance of the signature-based generation VAE(b) with the performance of returns-based generation VAE(a), see  \emph{Step 5} below, in particular the Section \emph{Performance evaluatio on the level of signatures}. In order to construct path from the generated VAE(a/i) returns, we take a Monte-Carlo-type approach by assembling daily generated increments to a full sample path.\smallskip
\subsubsection*{(Step 5) Evaluating the generated paths}
In this step we apply performance evaluation metrics to the output of the generator networks to evaluate certain relevant characteristics of the generated distribution whether they reflect the true distribution of the original data. These evaluation metrics are different from the objective function of the generative process itself. These final performance evaluation tests resemble the role of the Discriminator network in a GAN, with the difference that in our algorithm, performance evaluation is manual and only happens once.

\subsubsection{Distances for time-series and sample paths: A computationally efficient MMD metric for laws of stochastic processes.}\label{sec:MMDSignatures} 
When it comes to assessing the quality of a set of generated paths, being able to compute the distance between the laws of two stochastic processes becomes imperative. A naive metric based on the marginals of the two processes is bound to fail, as two stochastic processes can have identical marginals but very different laws. Instead, a metric that considers the entire law of the stochastic process is needed. Moreover, this metric has to be computable in order to make it practical. In \cite{CO18}, the authors propose a (computationally efficient) MMD for laws of stochastic processes based on signature kernels. As an application, they use this metric to develop a two-sample test for stochastic processes. In the context of this paper, this statistical test can be used to evaluate the quality of our market generator.
To asses whether a generative model is able to generate paths that are realistic with respect to a sample of real paths $Y_1, \ldots, Y_n$,  we sample from the generative model $n \in \mathbb{N}$, paths $X_1, \ldots, X_n$ and we apply the two-sample test proposed in \cite{CO18}. More specifically, we compute the signature-based MMD test statistic $T (X_1, \ldots, X_n; Y_1, \ldots, Y_n)$

\begin{align}\label{eq:MMDSig}
\begin{split}
&T (X_1, \ldots, X_n; Y_1, \ldots, Y_n) := \\
&\frac{1}{n(n-1)} \sum_{i, j; i \neq j} k(X_i, X_j) - \frac{2}{n^2} \sum_{i, j} k(X_i, Y_j) + \frac{1}{n(n-1)} \sum_{i, j; i\neq j} k(Y_i, Y_j),
\end{split}
\end{align}
where $k(\cdot, \cdot)$ is the so-called \textit{signature kernel} (see \cite[Proposition 4.2]{CO18}). Then, given a fixed confidence level $\alpha\in (0, 1)$, we compute the threshold $c_\alpha := 4\sqrt{-n^{-1} \log \alpha}$. The generative model will be said to be realistic with a confidence $\alpha$ if $T_U^2 < c_\alpha$.
\bigskip\\

\begin{center} \textbf{Performance evaluation on the level of signatures:} \vspace*{0.4cm}\\
\begin{tabular}{c||c|c|c|c|c}
\textbf{Gen. data$\downarrow$ / Real data$\rightarrow$}   & \textbf{Weekly paths (b/ii)} & \textbf{Monthly paths (b/iii)}\\
\hline 
\textbf{Returns (a/i,ii,iii)}  & Ret (a/i) $\Rightarrow$ paths $\Rightarrow$(b/ii)&  Ret (a/i) $\Rightarrow$ paths$\Rightarrow$(b/iii)\\
\hline 
\textbf{Log-Signatures (b/ii)}  & Direct & (b/ii)$\times4\Rightarrow$ Direct on (b/iii)\\
\hline 
\textbf{Log-Signatures (b/iii)}   & (b/ii)$\times4\Rightarrow$ Direct on (b/iii)& Direct\\
\hline 
\end{tabular}\\
\end{center}

%\red{Explain the above table in form of a paragraph instead of a table.}\\
\textbf{Comments:} 
 The vertical column denotes the generated data and the horizontal row the original data. In particular Signatures (b/ii) dentotes that the output of the generative model was on the level of weekly signatures. Therefore, to compare (b/ii) weekly generated signatures with (b/iii) signatures monthly paths, one concatenates four weekly signatures (see: product of signatures in Appendix \ref{sec:Signatures}) and compares on the monthly signature level.
 Clearly, if we generated data on the level of returns, we can compare this with the returns distribution of the original data direcly (see table below). But for a comparison on the level of signatures, as a first step one builds random paths, by sampling from the generated returns for each new increment, then calculates the corresponding signature of the thus generated paths. This process is encoded in the notation Ret (a/i) $\Rightarrow$ paths $\Rightarrow$(b/ii). \bigskip\\
 
\begin{center} \textbf{Performance evaluation on the level of returns / marginal distributions:} \vspace*{0.4cm}\\
\begin{tabular}{c||c|c|c|c|c}
\textbf{Gen. data$\downarrow$ / Real data$\rightarrow$} & \textbf{Daily data (a/i)} & \textbf{Weekly paths (a/ii)} & \textbf{Monthly paths (a/iii)}\\
\hline 
\textbf{Returns (a/i,ii,iii)} & Direct with (a/i) & Direct with (a/ii) &Direct with (a/iii)\\
\hline 
\textbf{Inverted Sig. (b/ii)} &  (b/ii)$\Rightarrow$paths$\Rightarrow$(a/i)&  (b/iii)$\Rightarrow$ paths$\Rightarrow$(a/ii)& (b/ii)$\Rightarrow$paths$\times 4$ $\Rightarrow$(a/iii)\\
\hline 
\textbf{Inverted Sig. (b/iii)} & (b/iii)$\Rightarrow$paths $\Rightarrow$ (a/i) &  (b/iii)$\Rightarrow$ paths $\Rightarrow$ (a/ii)&(b/iii) $\Rightarrow$ paths $\Rightarrow$ (a/iii)\\
\end{tabular}\\
\end{center}
%\red{Explain the above table in form of a paragraph instead of a table.}\\
\textbf{Comments:} If both the generated data is in form of returns (a/i,ii,iii), then comparisons of the empirical marginal distributions are direct. If, however, the data generation was on the level of signatures (b/ii,iii) then one first has to invert the signatures back to paths for a comparison. This is encoded in the entry (b/iii)$\Rightarrow$paths $\Rightarrow$ (a/i). Note that process of inverting signatures can be slow (the longer the paths the slower) and efficient algorithms for that are subject of onoing research, see for example \cite{C18}. 
\smallskip\\
\end{enumerate}

%\newpage
\section{Numerical Results}\label{sec:NumericalExperiments}

\subsection{Numerical experiments with historical data of S\&P}\label{sec:pathsrealisticimages} 
To demonstrate the accuracy generated log-signatures and images of the produced paths, we provide in this section images of 2D projections and the resulting generated paths and corresponding returns for optical demonstration. The rigorous numerical demonstration of the accuracy of the paths via the Maximum Mean Discrepancy inspired signature moments method can be found in Section \ref{sec:pathsrealisticMMD} below.

\begin{center}
\begin{figure}\label{fig:UncondSig}
\includegraphics[scale=0.35]{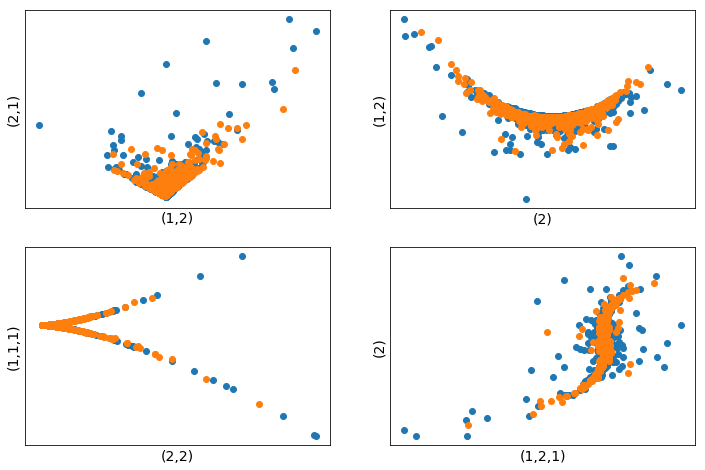}
%\vspace*{0.3cm}\\
\caption{\textbf{The (unconditional) VAE:} The image shows projections of generated weekly signatures (b/ii). Since the log-signatures we generate in this procedure are high-dimensional objects, we display here their projections on various two-dimensonal subspaces, indicated on the vertical axis.
}
\includegraphics[scale=0.35]{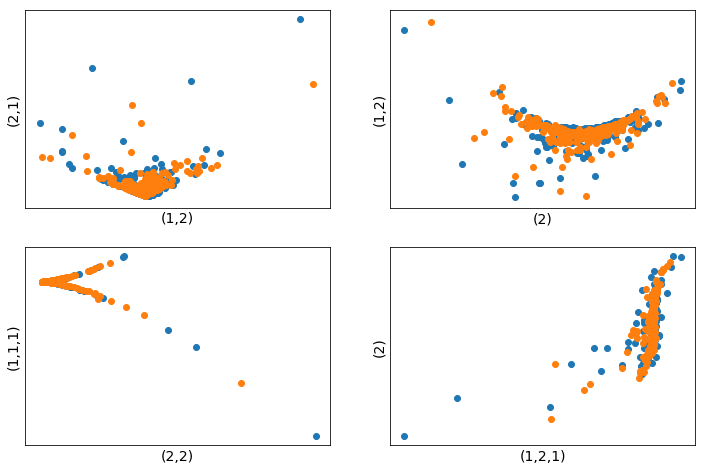}
\caption{\textbf{The (unconditional) VAE:} The image shows projections of generated monthly signatures (b/iii) on various two-dimensonal subspaces. \vspace*{0.3cm}}
\end{figure}
%%%%%%%%%%%%%%%%%%%%%%%%%
\begin{figure}\label{fig:Uncondpaths}
\includegraphics[scale=0.4]{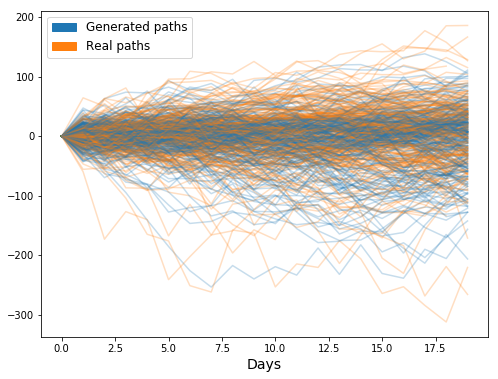}\includegraphics[scale=0.55]{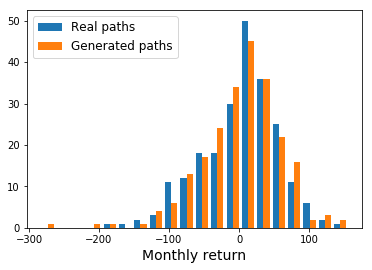}
\caption{This is the image of generated paths inverted from log-signatures.\vspace*{0.3cm}
}
\end{figure}
\end{center}
%%%%%%%%%%%%%%%%%%%%%%%%%

\begin{center}
\begin{figure}
\includegraphics[scale=0.48]{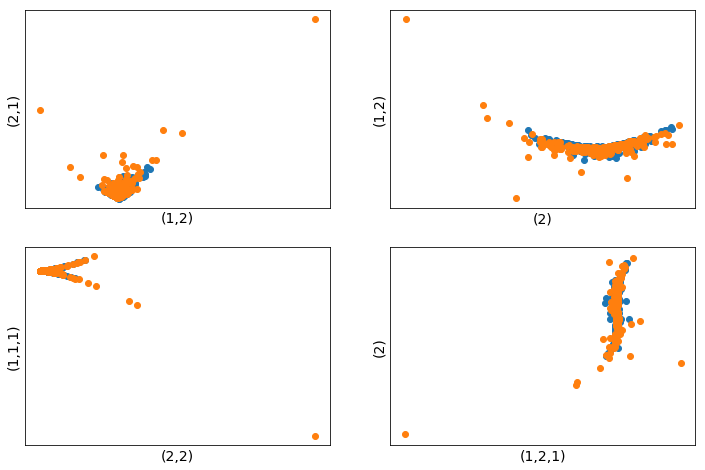}
%\vspace*{0.3cm}\\
\caption{\textbf{The conditional VAE:} The image shows projections of weekly signatures generated by the CVAE, projected on various 2dim subspaces. Here we condition on an instantaneous volatility of $5\%$ of the process measured a the start of the interval.}
\includegraphics[scale=0.48]{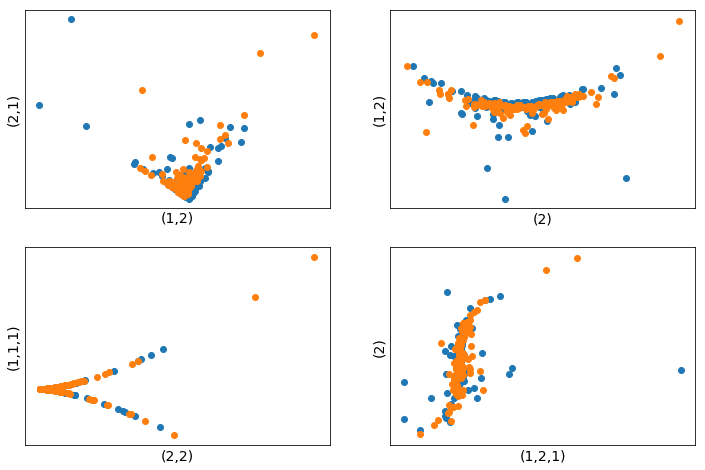}
\caption{\textbf{The conditional VAE:} The image shows projections of weekly signatures generated by the CVAE, projected on various 2dim subspaces. Here we condition on the current level of the process a the start of the interval. More specifically, we condition on the spot price being $S(0) =2000$. }
\end{figure}
\begin{figure}
%\includegraphics[scale=0.24]{Image1.png}\includegraphics[scale=0.24]{Image1.png}
%\caption{\red{Placeholder!} \textbf{The conditional VAE:} Projections of signatures generated by the CVAE, projected on various 2dim subspaces. Here we condition on the log-signature of the previous path. Weekly signatures (b/ii) in the left and monthly signatures (b/iii) in the right. }
\end{figure}
\end{center}
\newpage
\subsubsection{Performance evaluation scores}\label{sec:pathsrealisticMMD} We conduct the signature-based MMD two-sample test \eqref{eq:MMDSig} described in Section \ref{sec:MMDSignatures} for the scenarios described above. In the table below, we include 1.) the confidence level at which the the signature based MMD test changes from the result ``\emph{the two samples come from the same distribution}" to the result "\emph{the two samples come from different distributions}". Clearly, if the the test can be passed on a higher confidence level, it indicates a higher similarity of the generated samples to the original ones.\\
As a comparison we also performed a classical two sample test on the level of marginals, summarized in the table below: We include 2.) the corresponding confidence level for the statistic for the Kolmogorov-Smirnov test applied to the marginal distribution at final time of the time horizon.

We display the results for weekly and the monthly paths of the unconditional (uc) variational autoencoder, as well as one example for the conditional variational autoencoder, where we condition the samples of the instantaneous volatility being close to the $5\%$ level.\\
\begin{tabular}{c||c|c|c}
 &MMD signature confidence level& K-S test $p$-value\\
\hline
\hline
Weekly signature paths (uc)& $ 99.998 \%$ &0.57\\
\hline
Monthly signature paths (uc)& $99.987 \%$ &0.46\\
\hline
Weekly signature paths (conditional)& $99.997\% \%$ &0.63\\
\end{tabular}\\
\smallskip\\
Of course we could carry on here to include further statistics that control for properties of stylised facts from Section \ref{sec:Stylizedfacts} as it was done in previous works. However we recall here that the signature based MMD test of \cite{CO18} described in Section \ref{sec:MMDSignatures} completely characterizes the law of a stochastic process and therefore we omit further test measures here for brevity. \bigskip \\
Furthermore, below we also include the comparison of generated unconditional weekly paths that were learned as a 5-dimensional joint distribution of returns, which we compare with the weekly paths generated by the log-signature based generator. In the table below, we display  for these two sets of weekly generated samples (left side) the result of the signature-based MMD test at 99.95\% confidence level for the full path, as well as (right side) the result of the Kolmogorov-Smirnov test at 99.95\% confidence level applied to the daily marginal distributions from day 1 (denoted by K-S d1), to day 5  (denoted by K-S d5) of the week.\\ 
 \smallskip\\
 \begin{tabular}{c||c||c|c|c|c|c}
\textbf{At 99.95\% conf. level}&MMD signature test& K-S d1 & K-S d2&K-S d3&K-S d4&K-S d5\\
\hline
\hline
VAE  signatures & Passes & Passes & Passes & Passes & Passes & Passes\\
\hline
VAE multidim. distr.& \textbf{Fails} & Passes & Passes & \textbf{Fails} & \textbf{Fails} & Passes \\
\hline
\end{tabular}\\
\smallskip\\
We observe that the the latter greatly outperforms the former in terms of quality of generated paths with the same amount of training for both methods. 

\subsection{Numerical experiments with synthetic paths from rough volatility models}
Finally we demonstrate our experiments also on a fractional stochastic volatility model, the rough Bergomi model introduced in 2015 by Bayer, Friz and Gatheral \cite{BFG15}, which is a natural extension of the rough fractional stochastic volatility (RSFV) model in \cite{GJR14} to the setting of pricing. Rough stochastic volatility models have the advantage that they capture well some of the essential stylized facts of financial time-series.  
\begin{eqnarray*}
dX_t=&-\frac{1}{2}V_tdt+\sqrt{V_t}dW_t\qquad &X_0=\log(S_0)\\
V_t=&\xi_0 \ \mathcal{E}(2\nu C_H \mathcal{V}_t), \qquad &V_0, v, \xi_0>0\\
\mathcal{V}_t=&\int_0^t(t-u)^{H-1/2}dZ_u, \qquad &H\in(0,1/2)\\
\langle Z,W \rangle_t=&\rho t, \qquad &\rho \in (-,1,1),
\end{eqnarray*}
where $X:=\log(S)$, 
and $\mathcal{E}(\cdot)$ denotes the stochastic exponential and $C_H:=\frac{2H\Gamma (3/2-H)}{\Gamma(H+1/2)\Gamma(2-2H)}$. 
We include this experiment in order to demonstrate that our method already works optimally on the small number of paths available from S\&P 500 data. We demonstrate this by numerically generating a simulated training dataset from the rough Bergomi model with model parameters inspired by the calibration results in \cite{DLV19}. In the first experiment generate the same number of paths in the rough Bergomi model as the number of paths available from the S\&P training data (small dataset) and another training dataset with a significantly larger number of simulated paths (large dataset).We then train the Variational Autoencoder on the log-signatures of the simulated paths first on the small dataset and then on the large dataset and assess the quality of the generated paths by observing the value of the test statistic of the MMD two-sample test on the same confidence level for the two different outputs.
\begin{remark}
Since among classical stochastic volatility models rough volatility models are particularly realistic ones, we also assessed the quality of paths generated by Rough Fractional Stochastic Volatility comparing them with the original S\&P 500 monthly paths via the signature based MMD test statistic results, which can be found in the code accompanying this work, published on github. 
\end{remark}

\begin{center}
\begin{figure}
\includegraphics[scale=0.33]{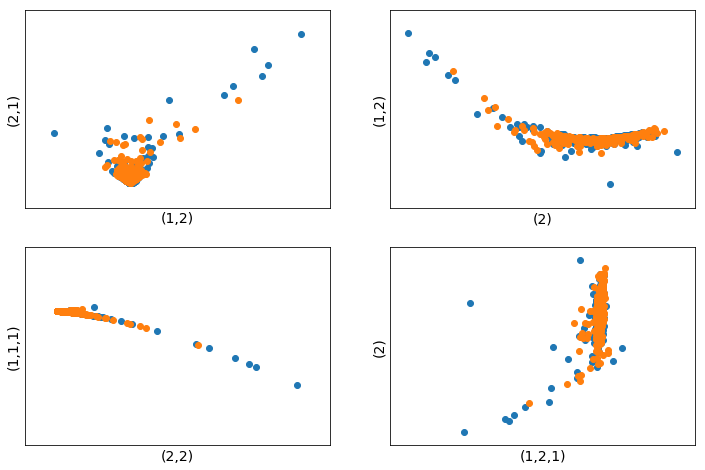}\includegraphics[scale=0.33]{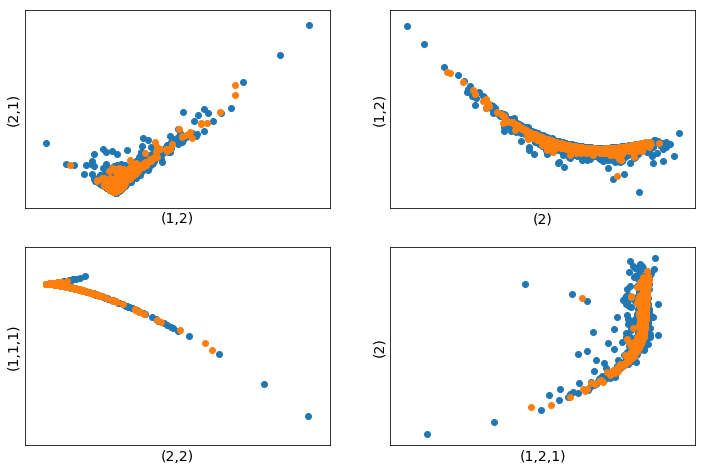}
\caption{Left had side: Projections of signatures generated by the VAE trained on 250 (small dataset) monthly rough Bergomi paths. Right had side: Projections of signatures generated by the VAE trained on 5000 (large dataset) monthly rough Bergomi paths.
}
\end{figure}
\end{center}
\section{Conclusions}
Our experiments show that the generative model works just as well but not significantly better if we have more original data in the training phase to calibrate the network parameters. Providing more training data does not significantly improve the learning process of the generative model. This demonstrates that our Variational Autoencoder based training is ideally suited for the scarce data environment at hand, operating efficiently with the little data that we have available for training (fewer than $\sim$ 250 samples). 
While returns-based optimisation works significantly better if it is provided with more training data (tested with numerically generated training data), the signatures-based training delivers convincing results already for the small amount of data from the S\&P paths and does not significantly improve if more (numerical) training samples are provided.

\begin{appendix}

\section{Signatures}\label{sec:Signatures}

\subsection{Signatures and their properties}\label{sec:signatureproperties}

In this section we recall properties of signatures that are used in this paper.
%\begin{definition}[Signature of a path]
%Let $X:[0, T]\to \mathbb R^d$ be a continuous path of bounded variation. The signature of $X$ is then defined by the sequence of iterated integrals given by
%$$\mathbb X_{T}^{<\infty} := (1, \mathbb X_{t}^1, \ldots, \mathbb X_{T}^n, \ldots)$$
%where
%$$\mathbb X^n_{T} := \int_{0<u_1<\ldots<u_k<T} dX_{u_1}\otimes \ldots \otimes dX_{u_k}\in (\mathbb R^d)^{\otimes n}$$
%with $\otimes$ the tensor product. Similarly, given $N\in \mathbb N$, the truncated signature of order $N$ is defined by
%$$\mathbb X_{T}^{\leq N} := (1, \mathbb X_{T}^1, \ldots, \mathbb X_{T}^N).$$
%\end{definition}
%
%\begin{remark}
%If the path $X$ has bounded variation -- which is the case of discrete data -- the integrals above can be defined using Riemann-Stieltjes integrals.
%\end{remark}
The signature of a path is a transformation of path space. Moreover, certain properties of signatures make them good feature sets for machine learning: 

\begin{theorem}[Uniqueness:\cite{horatio}]
Under certain assumptions (see \cite{horatio}) a path is uniquely determined by its signature.
\end{theorem}
Thus the signature map is a faithful transformation, in the sense that distinct paths have distinct signatures.
Therefore, signatures are feature maps that do not lose any information about the original path. Furthermore, for stochastic processes $X:[0,T]\to \mathbb R^d$, the expected signature will play a similar role to the moments of a random variable on a finite-dimensional vector space: under certain assumptions, it characterises its law.

\begin{theorem}[Expected signature characterises the law of a process:\cite{CL16}]
Let $X:[0,T]\to \mathbb R^d$ be a stochastic process on a probability space $(\Omega, \mathcal F, \mathbb P)$ such that its signature $\mathbb X_{0,T}^{<\infty}$  is a.s. well-defined. Assume that its expected signature $\mathbb E[\mathbb X_{T}^{<\infty}]$ is well-defined. Under certain assumptions (see \cite{CL16}) $\mathbb E[\mathbb X_{T}^{<\infty}]$ uniquely determines the law of the stochastic process $X$.
\end{theorem}
Hence, if one is interested in learning to generate samples from a certain stochastic process, one can instead learn to generate signatures of samples of the stochastic process. By the theorem above, this would be sufficient to completely characterise the law of the process, as the expected signature characterises the law. On the other hand, by the uniqueness of signature, knowing the signature of a path is essentially equivalent to knowing the path. This is precisely the approach we will follow in this paper: we will learn how to generate signatures of paths.\\

However, generating signatures directly is not an easy task because of the intrinsic structure of signatures. In other words, if a generative model is built to generate signatures directly, the generated object may not be the signature of any path. To avoid this, we will learn how to generate log-signatures instead:

\begin{definition}[Log-signature]\label{def:logsig}
Let $X:[0, T]\to \mathbb R^d$ be a path such that its signature $\mathbb X_{0,T}^{<\infty}$ is well-defined. The log-signature is then defined by
$$\log \mathbb X_{T}^{<\infty} := -\mathbb X_{T}^{<\infty} + \dfrac{1}{2}(\mathbb X_{T}^{<\infty})^{\otimes 2} - \dfrac{1}{3} (\mathbb X_{T}^{<\infty})^{\otimes 3} + \ldots + (-1)^n \dfrac{1}{n} (\mathbb X_{T}^{<\infty})^{\otimes n} + \ldots,$$
which can be shown to be well-defined (\cite{LTL07}).
\end{definition}

Taking the logarithm of the signature is an invertible operation -- one can exponentiate it to retrieve the signature. Therefore, no information is lost or gained when considering log-signatures. Log-signatures are defined on a certain free-Lie algebra and in fact, any element of this free-Lie algebra is the log-signature of a certain path (see \cite{LTL07} for more details). Hence, in this paper we will learn how to generate log-signatures of market paths, so that we can guarantee that the outputs of the ML generative model are indeed log-signatures.

\subsection{Lead-lag transformation}\label{sec:leadlag}

\begin{figure}
\includegraphics[width=\linewidth]{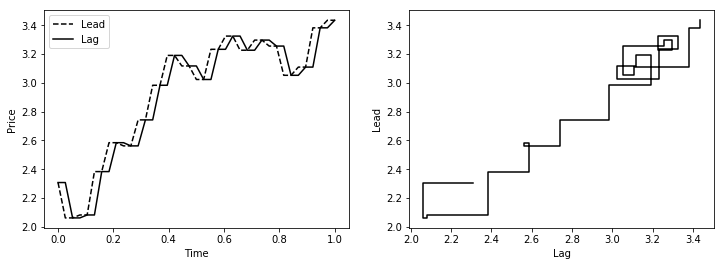}
\caption{Lead-lag transformation of a price path. The figure on the left shows the lead and lag components of the path, and the figure on the right shows the lag component plotted against the lead component.}
\label{fig:leadlag}
\end{figure}

In \cite{guy}, the authors introduce a transformation of path space called the \textit{lead-lag} or the \textit{Hoff} transformation. The authors showed that this transformation is able to capture the volatility of a path and, due to the importance of volatility in finance, we opted to use this transformation.

Let $D=\{t_i\}_{i=0}^n\subset [0,T]$, and let $\{X_{t_i}\}_{t_i\in D}\subset \mathbb R^d$ be a $d$-dimensional sample. The lead-lag transformation of $\{X_{t_i}\}_{t_i\in D}$ is defined below.

\begin{definition}[Lead-lag transformation]
The lead-lag transformation of $\{X_{t_i}\}_{t_i\in D}$ is defined by the $2d$-dimensional continuous path $X^D=(X^{D, b}, X^{D, f}):[0,T]\to \mathbb R^{2d}$ given by

\begin{align*}
X_t^D = (X^{D, b}, X^{D, f}):=  \begin{cases}
\left(X_{t_{k}},X_{t_{k+1}}\right), & \quad t\in\left [\frac{2k}{2nT},\frac{2k+1}{2nT}\right ),\\
\left(X_{t_{k}},X_{t_{k+1}}+2(t-(2k+1))\left(X_{t_{k+2}}-X_{t_{k+1}}\right)\right), & \quad t\in\left [\frac{2k+1}{2nT},\frac{2k+\frac{3}{2}}{2nT}\right ),\\
\left(X_{t_{k}}+2(t-(2k+\frac{3}{2}))\left(X_{t_{k+1}}-X_{t_{k}}\right),X_{t_{k+2}}\right), & \quad t\in\left [\frac{2k+\frac{3}{2}}{2nT},\frac{2k+2}{2nT}\right ).
\end{cases}
\end{align*}

The component $X^{D, b}$ is the backward or lag component, and $X^{D, f}$ is the forward or lead component. The signature of the lead-lag transformation will be denoted by $\mathbb X_{0,T}^{D, <\infty}$.
\end{definition}

Figure \ref{fig:leadlag} shows the lead-lag transformation of a certain path. As the name suggests, the lead component is \textit{leading} the lag component. As it was shown in \cite{guy}, the relationship between the lead and lag component is able to capture the volatility of the path. For instance, if the sample $\{X_{t_i}\}_{t_i\in D}\subset \mathbb R^d$ comes from a $d$-dimensional continuous semimartingale, when the size of the mesh tends to 0 the authors in \cite{guy} showed that $\mathbb X_{0,T}^{D, <\infty}$ converges to a certain rough path that incorporates information about the quadratic variation -- i.e. volatility -- of the semimartingale.

\section{Variational Autoencoders}\label{sec:VAE}
In this section we briefly recall the basics of Variational Autoencoders (henceforth VAEs), our choice of generative model. More details and background information on VAEs can be found in \cite{MLBook, KingmaVAE, KingmaVAE2}.  VAEs have been recently itroduced in the pioneering 2014 article of Kingma and Welling \cite{KingmaVAE}. A particularly applealing feature of that work is, as they emphasize, that their results can by construction be applied to nonstationary settings, such as time-series data \cite[Section 2]{KingmaVAE}. 
In the section below, we lay out also further reasons of our motivation for choosing VAEs as a generative model for our Market Generator. The basic mechanism of variational autoencoder essentially rely on a Maximum Likelihood idea (see \cite{%Doer16, 
MLBook}), adjusting the  generative process via backpropagation to maximize (lower bound of) the probability of observing the given training samples. \begin{align}\label{eq:VAEmaximization}
P(X) = \int P(X|z;\theta) P(z) dz
\end{align}
\noindent 
It is common practice to choose the initial distribution as a d-dimensional Gaussian.\\
\begin{align}\begin{split}\label{eq:VAENormal}
P(X|z;\theta)&=\mathcal{N}(f(z;\theta, \sigma^2*I)), \quad \textrm{where $\sigma$ is to be set as a hyperparamter}\\
P(z)&=\mathcal{N}(z|0,I)
\end{split}
\end{align}
The basic functioning of a VAE is the following: Given one random variable $z$ with one distribution, we can create another random variable  $X= g(z)$ with very different distribution: The deterministic function $g$ is then learned from the data through the function approximation capacity of the neural network.\\
If using these expressions \eqref{eq:VAENormal}, we can find a (continuously differentiable) expression for $P(X)$ in \eqref{eq:VAEmaximization}, then we can optimize the model using stochastic gradient descent to update the network parameters $\theta$.
By gradient descent, we gradually make the training data more probable under the generative model.
%\blue{Clearly at this point we should note that the underlying assumption here is that the given data sample is representative of the underlying distribution.}
The equation \eqref{eq:VAEmaximization} implies two tasks for VAEs to solve (i) how to define the latent variable space $\mathcal{Z}$ (what information they represent and what is the structure between its dimensions) making sure that the latent variables capture the relevant information in the generative process and (ii) taking the integral over $z$ in \eqref{eq:VAEmaximization}.
Regarding problem (i) VAEs avoid explicitly describing the dependencies between dimensions of $z$, assume no latent structure, instead, they rely on the inherent property of neural networks as functional approximators.  
Finally, it is worth mentioning that Variational Autoencoders are called ``autoencoders'' because the training objective \eqref{eq:VAEmaximization} that derives from this setup has an encoder and decoder structure that resembles a traditional autoencoder, the non-linear generalisation of PCA. For more details and background on Variational Autoencoders see the original work \cite{KingmaVAE} and for a gentle introduction see \cite{KingmaVAE2%, Doer16
}.\\

\end{appendix}
%%%%%%%%%%%%%%%%%%%%%%%%%%%%%%%%%%%%%%%%%%%%%%%%%%
%%%%%%%%%%%%%%%%%%%%%%%%%%%%%%%%%%%%%%%%%%%%%%%%%%

\end{document}